\documentclass[12pt,a4paper]{article}
\usepackage{a4wide}
\usepackage{latexsym}
\usepackage{epsf}
\usepackage{amssymb}
\begin{document}
\def\be{\begin{equation}}
\def\ee{\end{equation}}
\def\bea{\begin{eqnarray}}
\def\eea{\end{eqnarray}}

\def\pd{\partial}
\def\a{\alpha}
\def\b{\beta}
\def\g{\gamma}
\def\d{\delta}
\def\m{\mu}
\def\n{\nu}
\def\t{\tau} 
\def\l{\lambda}
\def\s{\sigma}
\def\e{\epsilon}
\def\scri{\mathcal{J}}
\def\cM{\mathcal{M}}
\def\tcM{\tilde{\mathcal{M}}}
\def\RR{\mathbb{R}}
\def\CC{\mathbb{C}}

\hyphenation{re-pa-ra-me-tri-za-tion}
\hyphenation{trans-for-ma-tions}


\begin{flushright}
IFT-UAM/CSIC-99-28\\
hep-th/9907205\\
\end{flushright}

\vspace{1cm}

\begin{center}

{\bf\Large Ultraviolet and Infrared Freedom from String Amplitudes}

\vspace{.5cm}

{\bf Enrique \'Alvarez}${}^{\diamondsuit,\clubsuit}$
\footnote{E-mail: {\tt enrique.alvarez@uam.es}}
{\bf and C\'esar G\'omez}${}^{\diamondsuit,\spadesuit}$
\footnote{E-mail: {\tt iffgomez@roca.csic.es}} \\
\vspace{.3cm}

\vskip 0.4cm

${}^{\diamondsuit}$\ {\it  Theory Division, CERN,1211 Geneva 23, 
Switzerland \\and\\ 
Instituto de F\'{\i}sica Te\'orica, C-XVI,
  Universidad Aut\'onoma de Madrid \\
  E-28049-Madrid, Spain}\footnote{Unidad de Investigaci\'on Asociada
  al Centro de F\'{\i}sica Miguel Catal\'an (C.S.I.C.)}

\vskip 0.2cm

${}^{\clubsuit}$\ {\it Departamento de F\'{\i}sica Te\'orica, C-XI,
  Universidad Aut\'onoma de Madrid \\
  E-28049-Madrid, Spain }

\vskip 0.2cm

${}^{\spadesuit}$\ {\it I.M.A.F.F., C.S.I.C., Calle de Serrano 113\\ 
E-28006-Madrid, Spain}

\vskip 1cm


{\bf Abstract}

The infinite tension limit of string amplitudes is examined with some care,
identifying the part responsible of diamagnetic behaviour as well as
a peculiar paramagnetic {\em tachyon magnifying} responsible of
 aysmptotic freedom. The way string theory represents abelian gauge theories
is connected with the non-planar reggeon/pomeron amplitude and a nontrivial
beta function is found in the low energy limit of a single D-brane.

\end{center}

\begin{quote}

\end{quote}


\newpage

\setcounter{page}{1}
\setcounter{footnote}{1}

\section{Introduction}
One of the most important results in quantum field theory 
was the discovery \cite{gross} of asymptotic freedom 
for non abelian gauge groups.
String theory, among other things, can be used as a new 
regulator of quantum field theories with the ultraviolet cutoff
given by the string tension $\alpha'$. Appropiated field theory
limits, $\alpha'=0$, can be defined for string amplitudes and there
is now an impressive body of knowledge (cf. for example,\cite{metsaev},
\cite{minahan}.
\cite{divecchia},
\cite{bern}, and references therein) on the recovering,  in this limit, of 
the perturbative series of non abelian gauge theories and in particular
the correct asymptotically free beta function.
\par
At the level of open string amplitudes the difference between dealing
with abelian or non abelian gauge groups reduces simply to the type of 
Chan-Paton factors used to decorate the end points of the open string.
These group theoretical elements appear in string theory as
multiplicative factors of the skeleton string amplitude that are independent
of the gauge group. Thus from the string point of view
the only difference between abelian and non abelian
gauge theories stems basically from the existence of a no planar contribution
to the two point amplitude in the abelian case that vanishes 
for non abelian Chan-Paton factors. 
Therefore and using string theory as the driving principle we notice
that the appearance of infrared free theories is intimately related to
the existence of the old fashion reggeon-pomeron amplitude on the basis 
of which we build up the the two point non planar string diagram.
\par
Our first aim in this paper is to perform a careful analysis, in
the infinite tension limit, of some open string two point 
functions and to unravel
the specific way string theory produces infrared free beta functions.
From this analysis we shall get some interesting physical results.
\par
First of all we observe that string theory produces a decomposition
of the beta function into two pieces, infrared free and asymptotically free
respectively, corresponding exactly to the diamagnetic
and paramagnetic contributions to the pure Yang Mills beta function 
(\cite{nielsen}). 
When we take into account the non planar two point function -specific
to abelian Chan-Paton factors- we cancel the paramagnetic asymptotically
free part ending up with a non trivial infrared free beta function
corresponding to scalar quantum electrodynamics (QED) with
 $(d-2)$ charged scalar fields, in $d$ 
 space-time dimensions. We then observe that the way strings, 
manage to produce infrared free behavior is
through the non planar (reggeon/pomeron vertex)  contribution
which precisely cancels the paramagnetic (spin-dependent) 
piece of the pure Yang Mills 
beta function, while keeping untouched the diamagnetic part
which shows up eventually as associated with a set of 
unexpected charged scalar fields.
\par
In modern terminology Chan-Paton factors become equivalent to D-branes and the
difference between non abelian and abelian equivalent to that between 
a family of parallel D(irichlet)-branes and one isolated D-brane.
\par
Using D-3 branes we can repeat our analysis for the beta function
in the infinite tension limit, but this time for the open string amplitude
with world sheet boundaries located on the D-3 brane world volume. For the
non abelian case we get -at one loop- the expected result, namely
the Yang Mills beta function for a theory with $(D-4)$ charged scalars
in the adjoint representation (where $D$ is the external spacetime 
dimension). It is interesting to notice that this beta
function precisely vanishes for $D$ the bosonic string critical dimension $26$.
Moreover the so derived beta function agrees with the expected
spectrum on a set of parallel D-3 branes in the bosonic string, namely
Yang Mills gluons and adjoint charged scalars, one for each
transverse direction (the Goldstone bosons of the
spontaneously broken translation invariance).
\par
For the case of one D-brane and once we take into account the non planar
contribution to the two point function, we get a beta function for scalar
QED with $(D-2)$ charged scalars, a matter content with an unclear
geometrical interpretation. In any case we notice that what will make
the world volume dynamics on one D-brane infrared free is intimately
related to the non vanishing amplitude for spontaneous
emission processes of the type depicted in figure 1.

\begin{figure}[!ht] 
\begin{center} 
  \leavevmode \epsfxsize= 10cm \epsffile{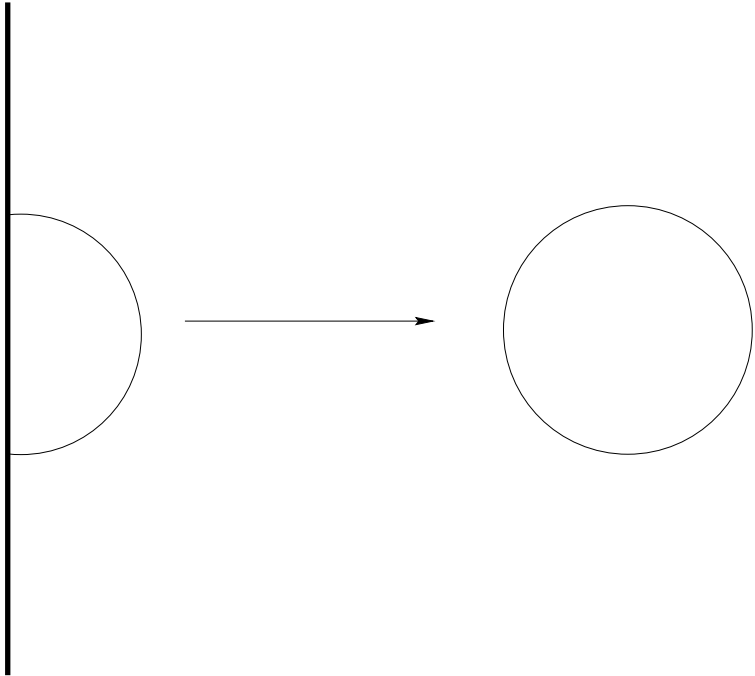}
\caption{ Open/closed string transition from D-branes to the bulk}
\label{fig:uno} 
\end{center} 
\end{figure}

If by some stability argument we can freeze this emission we will end up
with an asymptotically free dynamics on the world volume of the D-brane.

\section{Field Theory Limit of String Amplitudes}
String techniques have been successfully used in the last decade
to reproduce field theory amplitudes in a way leading to further
insights in the computation of field theory Feynman diagrams.
(cf. \cite{bern}).
\par
In order to be specific, let us start by reviewing the {\em recipe of
the field theory
limit} in the simplest setting aiming to deriving the $SU(N)$ gluon
propagator from the open bosonic string (cf. \cite{divecchia}).
\par
This amplitude is given by 
\be C_{ab}(k_1,k_2)\equiv Tr
<\epsilon_1^{\m}V_{a \m}(k_1)\epsilon_2^{\n}V_{b \n}(k_2)> 
\ee 
where
the gauge vector vertex operator is given by \be V_a^{\m}(k)\equiv
g_{YM} \sqrt{2\a'} T_a\frac{dx^{\m}}{d\lambda} e^{i k.X(\lambda)} \ee
the matrices $T_a$ being the generators of the gauge group $SU(N)$.
The parameter $\lambda$ indicates the insertion point on the boundary of
the world surface of the string, and has to be integrated over. The
 Yang-Mills coupling constant is related to the string coupling, $g_s$,
through the D-brane relationship $g_{YM}=2(2\a')^{\frac{d-4}{4}}
g_s\equiv 2^{\frac{d}{4}}l_s^{\frac{d-4}{2}}g_s $.The
simplest one loop contribution is the annulus (which is topologically
a cylinder), which we shall parametrize, following \cite{green} by the
annular region shown in the figure, with the two arcs identified in
the sense shown by the arrows. (This is gotten by defining
the convenient variables $\omega\equiv \lambda_1 \lambda_2\in(0,1)$ and 
$\rho\equiv \lambda_1\in(\omega,1)$).
\par
The modulus of the cylinder is $\tau\equiv - \frac{1}{2}\log ~\omega 
\in (0,\infty)$
(whereas $\omega\in (0,1)$). Instead of the insertion
point on the boundary $\rho\in (\omega,1)$, we will use the convenient
 variable $\n \equiv - \frac{1}{2}\log ~\rho$.
\par
The diagrams can be classified as planar and non-planar, 
depending on whether
the two vertex operators are inserted on the same boundary 
(planar) or in different
boundaries (non planar).
We normalize the generators by $tr T_a T_b \equiv C^{fund}_2(G)
 \delta_{ab}=\frac{1}{2}\delta_{ab}$. 
\par
The infinite tension limit $\a'\rightarrow 0$ is expected to be dominated
by configurations in which $\tau\rightarrow\infty$.
This corresponds to cylinders with divergent section (that is, infinite time
from the open string point of view, and  only the massless 
open string states can survive in this limit).This is then the natural
{\em infrared} limit from the open string point of view.
\par
Owing to conformal invariance, the situation depicted is equivalent
to a cylinder with constant section, and length $l=\frac{1}{\tau}$. This would
be the most natural way to look at the cylinder from the closed string
channel point of view. The limit 
we are taking is then the limit in which the length of this cylinder goes
to zero.

\begin{figure}[!ht] 
\begin{center} 
  \leavevmode \epsfxsize= 10cm \epsffile{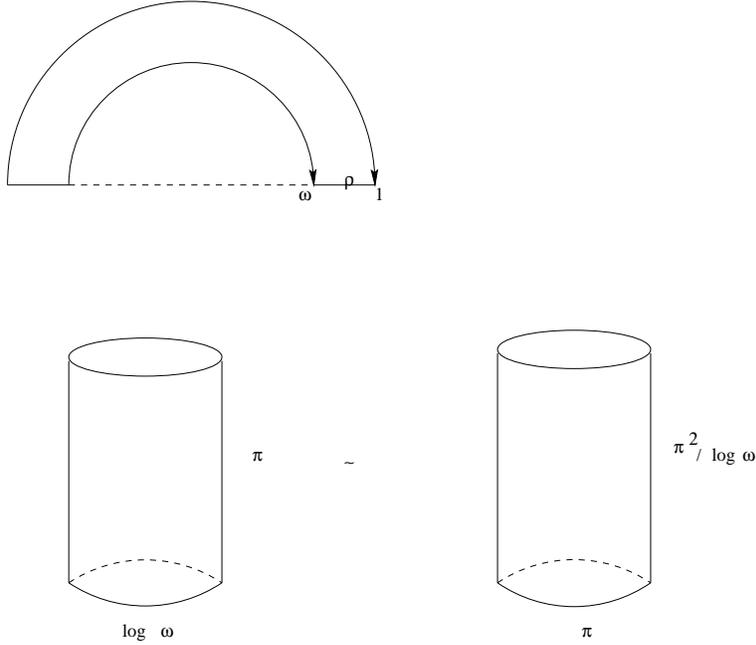}
\caption{ The annular region of integration for the one-loop diagram}
\label{fig:dos} 
\end{center} 
\end{figure}

The planar amplitude (which gives the full contribution of the simple
group $G=SU(N)$ under consideretion here) is given by(\cite{green})
\footnote{We follow the explicit
normalization of Di Vecchia et al. 
in \cite{divecchia}; including in particular 
their prescription for an off-shell continuation, namely, to gauge-fix
the projective transformations on the Koba-Nielsen variables and to continue
the momenta to $k_i^2 = M^2$, while keeping transversality, 
$\epsilon_i k_i = 0~\forall ~i$ . This prescription is {\em different} 
from the one advocated in \cite{bern}, but it is much simpler for our
purposes, and it has not led to any known inconsistencies.}:
\bea\label{amplii}
&& C_{ab}(k_1,k_2)= N\frac{g^2\mu^{4-d}}{(4\pi)^{d/2}}\frac{1}{2}\delta_{ab}
(2\a')^{1-d/2}
\nonumber\\
&&\int_{0}^{\infty}d\tau e^{\frac{(26-d)\tau}{12}}
\eta(i\tau/\pi)^{-(d-2)}\int_{0}^{\tau}d\nu \tau^{-d/2} 
\nonumber\\
&&[-(\epsilon_1.\epsilon_2) \pd_{\nu_1}\pd_{\nu_{2}}G(\nu_1,\nu_2) + 
\a'(\epsilon_1.p_2)(\epsilon_2.p_1)\pd_{\nu_1}
G(\nu_1,\nu_2)\pd_{\nu_2}G(\nu_1,\nu_2)]\nonumber\\
&&e^{2\a' p_1.p_2 G(\nu_1,\nu_2)}
\eea
$G(\nu_1,\nu_2)$ is the standard \footnote{This Green function is 
$2\pi$ times the one used in \cite{burgess}
\be
G_{BM}(z|\tau)\equiv \log~ |\frac{\theta_1(z|\tau)}{\theta'_{1}(0|\tau)}| - 
\pi~\frac{(Im~z)^2}{Im ~\tau} 
\ee
with the important difference that it does not have modulus in the logarithm.
Note in particular that $G_{BM}(z+\tau|\tau) = G_{BM}(z|\tau)$, whereas
$G(\nu+\tau|\tau)= G(\nu|\tau)+ i \pi $}
Neumann Green function (cf.\cite{green}),
\be
G(\nu|\tau)\equiv \log~- 2\pi i ~\frac{\theta_1(i\frac{\nu}{\tau}|i
\frac{\tau}{\pi})}
{\theta'_{1}(0|i\frac{\tau}{\pi})} - 
~\frac{\nu^2}{\tau} 
\ee
\par
There are many points of interest in this formula. 
The traces over the Chan-Paton on the 
two boundaries give $tr~1=N$ for the boundary free
of external states and $tr T_aT_b = \frac{1}{2}\delta_{ab}$ for the
boundary where the vertex operators lie.
\par
The amplitude has been boldly
written  in arbitrary dimension $d$, in spite of the fact that
we know that the string theory is only unitary if $d=26$. We will eventually
be intested in $d=4$, and we are expressing the coupling constant
in terms of the {\em dimensionless} four-dimensional one,
so that we  have included a term
$\mu^{4-d}$.
\par
A further point is that, {\em on shell}, by
 momentum conservation, $p_1=-p_2$, so that
$p_1^2=p_2^2=p_1.p_2 = p_1\epsilon_1=p_2\epsilon_2=0$, and the only surviving 
term is the one involving a second
derivative of the Green's function (which gives zero upon integration).
\par
The standard procedure (see \cite{bern}\cite{divecchia}) is to
integrate by parts precisely this term, and stay with the rest, making
analytic continuation in the Mandelstam variable 
\be 
s\equiv p_1.p_2 
\ee 
In order to take the infinite tension limit of the amplitudes, we shall
take simultaneously $\a',\t^{-1}\rightarrow \infty$, un such a way that
the combination $\bar{\t}\equiv \a'\t$ stays finite. Actually, this same
variable $\bar{\t}$ will play the r\^ole of Schwinger's proper time
in the field theory limit. Let us stress that due to the conformal mapping
depicted in Figure 2, the  $\t^{-1}$ is the {\em world sheet length}
of the cylinder, so that the limit $\t\rightarrow\infty$ corresponds to
 zero length where the dominant contribution corresponds to the 
lightest open string states.
\par
In this limit, the contribution of the partition function (conveyed by
Dedekind's function) is given by:

\be\label{dedekind} 
1 + (d-2)e^{-2\tau} 
\ee 
corresponding to the open tachyon and gluon contribution to the loop.
Once we multiply (\ref{dedekind}) by the measure factor 
$ e^{2\t}$ we get
the standard tachyon divergence as well as a finite contribution
for the gluonic piece in (\ref{dedekind}).
The terms involving the Green's function can be represented by:

\be\label{R}
R(s)=\frac{1}{s}
\int_0^{\infty}\frac{d\tau}{\tau}[e^{2\t} + (d-2)]
\int_0^{\tau} d\nu \tau^{1-d/2} e^{2\a's G(\nu)}
(\pd_{\nu}G(\nu))^2 
\ee

\subsection{The Tachyon  magnification }
The large $\tau$ expansion of the Green function is:
\be\label{green1}
G(\nu)= -\frac{\nu^2}{\tau} + \nu  + \log (1 - e^{-2\nu}) +
e^{-2\tau}[ 3 + \frac{e^{-3\nu}- e^{3\nu}}{e^{\nu}-e^{-\nu}}]
\ee
which can as well be written as:
\be\label{green2}
G(\nu)= -\frac{\nu^2}{\tau} + \nu  -\sum_{n=1}^{\infty}\frac{e^{-2n\nu}}{n}
+ e^{-2\tau}[3 - \sum_{n=0}^{\infty}e^{2(1-n)\nu}+\sum_{n=0}^{\infty}
e^{-2(n+2)\nu}]
\ee

Let us now consider the region $\n = \t \rightarrow\infty$ as defining a
 new integration variable $\hat{\n}\equiv \n/\t$. The relevant contribution
to the integrand of (\ref{R}) is given by:
\be
[e^{2\tau}+ (D-2)] (\pd G)^2=(D-2)(1-2\hat{\nu}^2) + 4 e^{-4\hat{\nu}\tau}
+4 (1-2\hat{\nu})e^{-2\hat{\nu}\tau} -8 + 4 e^{(4\hat{\nu}-2)\tau} - 4 
(1-2\hat{\nu})e^{2\hat{\nu}\tau}
\ee
The correct procedure in this limit, according to the above rules
is to neglect the two last terms of the above expansion, and then, perform
the integration over $d\hat{\nu}$ of the remaining integrand. 
Let us insist that
we have to neglect the term $4 e^{(4\hat{\nu}-2)\tau} $, 
in spite of the fact that
it enjoys a finite limit when
$\tau\rightarrow\infty$ as long as  $\hat{\nu}\leq 1/2$. This means that
it is necessary to neglect as tachyon artifacts all terms that diverge
when $\tau\rightarrow\infty$ for {\em some} value of $\hat{\n}$.
\par
The simplest remaining term, proportional to $(d-2)$, comes from
the {\em zero mode} part of the Green function, and is given by:
\be
e^{2\a' s (\hat{\nu}-\hat{\nu}^2)\tau} (d-2)(1-2\hat{\nu})^2 
\ee 
The remaining contribution, which can be interpreted as due to some sort
of {\em tachyon magnification} (because there is a cancellation of
the divergent contribution of the tachyon with an exponentially suppressed
contribution coming from oscillatory modes in the Green function), just
gives:
\be
e^{2\a' s (\hat{\nu}-\hat{\nu}^2)\tau} [ - 8]
\ee 
It should be stressed that by working in the combined limit
$\a',\t^{-1},\nu^{-1}\rightarrow\infty$, with the {\em blow-up}
variables $\tilde{\t}\equiv\a' \t$ and $\hat{\n}\equiv \n/t$, we have
 avoided the singular region in moduli space corresponding to coincident
vertex operators (cf. Figure 3), which from the field theory
point of view corresponds to the two last diagrams of Figure 4, which
do not contribute to the propagator.
\par

\begin{figure}[!ht] 
\begin{center} 
  \leavevmode \epsfxsize=10cm \epsffile{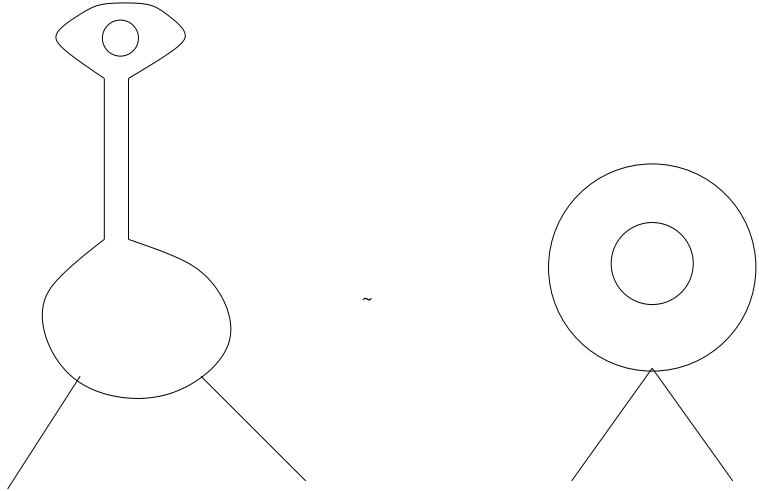}
\caption{ Two conformally equivalent ways of depicting coincident vertex 
operators}
\label{fig:tres} 
\end{center} 
\end{figure}

All this yields
\be
R(s)= - \Gamma(2-d/2)(-2\a's)^{d/2-2}\frac{7d-6}{d-1}B(d/2-1,d/2-1)
\ee
(where $B(z,w)$ is Euler's beta function).
It is useful for later purposes to realize that
\be\label{desc}
\frac{7d-6}{d-1}= 8 -\frac{d-2}{d-1}.
\ee

\begin{figure}[!ht] 
\begin{center} 
  \leavevmode \epsfxsize=10cm \epsffile{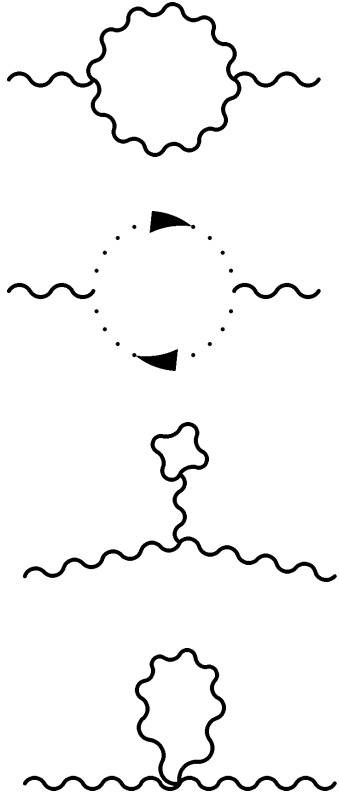}
\caption{  Diagrams for the Gluon Propagator. The wavy line represents gluons;
the discontinuous one, ghosts.}
\label{fig:cuatro} 
\end{center} 
\end{figure}

The final result of the (divergent part of the )two-gluon 
amplitude, computed in this way, is:
\bea\label{ampli}
&&C_{ab}(s)= \frac{N}{2}\d_{ab} \frac{g^2}{(4\pi)^{d/2}}(2\a')^{2-~d/2}
(\e_1.\e_2)~s~R(s) = \nonumber\\ 
&&- \frac{N}{2} ~\delta_{ab}~\frac{g^2}{(4\pi)^2}
~\epsilon_1.\epsilon_2 ~s ~\Gamma(\epsilon)~\frac{22}{3}
~B(1,1)
\eea
(where in the second line dimensional regularization around $d=4$
has been used).
It is remarkable that this is {\em exactly} the background field 
Feynman gauge result.
Indeed, using the Ward identity $Z_g= Z_A^{-1/2}$, one gets
\be
g_0\equiv \mu^{\epsilon}\sum_{n=0}^{\infty}\frac{a_n(g)}{\epsilon^n}=g\mu^{\epsilon}
(1 - N \frac{g^2}{32(\pi)^2}\frac{11}{3}\frac{1}{\epsilon}) + 
0(\frac{1}{\epsilon^2})
\ee
This leads to the correct gauge theory $\beta$ function, namely:
\be
\beta(g)\equiv\mu\frac{\pd g}{\pd \mu}= -a_1 + g\frac{\pd}{\pd g} a_1 =
 - \frac{11}{3}C_2^{adj}(G)\frac{g^3}{16\pi^2}
\ee
(where $C_2^{adj}(G=SU(N))~=~N$).
\par
 Indeed, the $\frac{11}{3} = 4 - \frac{1}{3}$ \footnote{This 
decomposition was previously found
(in the form $(D-26)/24 = (D-2)/24 -1$) in \cite{metsaev}) when they were
looking for the string theory effective action.}
, where the infrared free part
comes from the $(d-2)e^{-2 \tau}$ part in the expansion 
of the Dedekind function, and 
the {\em whole} ultraviolet free part comes from the $1$ (that is, 
the tachyon magnification).\par
 This whole setup
is to be contrasted with the corresponding calculation in QFT, where
both the contributions of the ghosts and the vector bosons are UV free
(11/3 = 10/3 + 1/3). 
\par
\subsection{The r\^ole of the off-shell extension}
It has already been pointed out that the amplitude (\ref{ampli})
is zero on shell, so that it had to be continued off-shell to get
a non-vanishing result. This step is a perfectly banal one in field theory
but full of dangers instead in string theory, owing to the fact that
only  on shell are the vertex operators conformal fields of scale dimension
$(1,1)$, which is in turn necessary for consistency.
\par
In the language of the operator formalism (\cite{alvarezgaume}), the on-shell
condition on the external states is necessary to prove that physical 
amplitudes are independent of the local coordinates of the punctures
corresponding to the insertions of the vertex operators. When the amplitude
is extended off-shell, changes of the coordinates of the punctures
do change the physical amplitudes. In reference (\cite{divecchia}) it has
been argued that in the field theory limit defined as above, those
changes are equivalent to changes of gauge fixing. 
\par
The fact that we recover in this way the background field Feynman gauge
(with the associated Ward identity) explains why we were able to extract
the beta function from the gluon propagator alone. Although we will have
no use for it, it is also true that tree and loop amplitudes are naturally 
recovered in different gauges, another peculiar property of the 
background field (cf. the second reference in \cite{abbott}).

\section{Stringy interpretation of asymptotic freedom}

It would be very interesting to find a field theory interpretation
of the natural string decomposition of the beta function in a
infrared free part (coming fromn the zero modes) and the asymptotically 
free piece, visible only thanks to the tachyon magnification.
It appears that this decomposition naturally fits a beautiful heuristic
argument attributed to 't Hooft by N.K. Nielsen (cf. \cite{nielsen}), 
where the two pieces are
called dia and paramagnetic because of the striking similarity 
of the gauge theory vacuum with a polarizable medium.
\par
Let us start by expanding the gauge field around a background $\bar{A}_{\m}$:
\be
A_{\m}=\bar{A}_{\m} + a_{\m}
\ee
so that
\be
F_{\m\n}=\bar{F}{\m\n} + \bar{\nabla}_{\m} a_{\n}-\bar{\nabla}_{\n} a_{\m}
+ [a_{\m},a_{\n}]
\ee
The Yang-Mills Lagrangian then reads:
\bea\label{bf}
&&L = -\frac{1}{4}\bar{F}_{\m\n}\bar{F}^{\m\n}  - \frac{1}{2}\bar{F}_{\m\n}
( \bar{\nabla}_{\m} a_{\n}-\bar{\nabla}_{\n} a_{\m})
-\frac{1}{2} (\bar{\nabla}_{\m}a_{\n})^2
-\frac{1}{2}\bar{F}^{\m\n}[a_{\m},a_{\n}]\nonumber\\
&&+\frac{1}{2}(\bar{\nabla}_{\m}a_{\m})^2
-\frac{1}{2}(\bar{\nabla}_{\m} a_{\n}-\bar{\nabla}_{\n} a_{\m})[a_{\m},a_{\n}]
\eea
The terms linear in the quantum fields vanish if the background field 
is on shell. On the other hand, up to the one loop approximation, only terms
quadratic in the quantum fields do contribute. Now it is clear that the 
{\em diamagnetic} (i.e., infrared free) part
will come from the covariant derivative term; actually, the current in the
loop induces, via the well-known Lenz's law, a magnetic moment which tends
to cancel the external magnetic field; this leads to the screening of the
charge or what is the same, in QFT language, to infrared freedom.
\par
After including the effects of the gauge fixing $\bar{\nabla}_{\m}a_{\m}=0$,
 embodied in the ghost fields,
this part of the amplitude (up to multiplicative group theory factors) goes as
the vacuum polarization in scalar QED with $d-2$ scalars (the true physical
polarizations). This is {\em exactly} the contribution we get in string
theory coming from the {\em zero modes} which  corresponds to
 the $1/3$ in the former 
units.
\par
The {\em paramagnetic} term, coming from the spin dependent interaction
(the last term in equation (\ref{bf})) (and where, we insist, {\em all}
the asymptotic freedom  resides) matches correspondingly the tachyon
magnification term from the stringy viewpoint that is, the $-4$ in the
 same units
as before. It is hard to believe that
 this is only a coincidence; everything points in the direction that there is
still much to be learned on the physical meaning of asymptotic freedom from
the study of string amplitudes.

\section{Physical effect of D-branes}
We have considered up to now formal string amplitudes in $d=4$
dimensions. In spite of the unavoidable off-shell continuation, in
the limit $\tau\rightarrow\infty$ we have encountered no problems with
unitarity, nor any other inconsistency.
\par
There is, however another possibility, namely to work with {\em critical}
strings (i.e., $D=26$), and still consider gluonic amplitudes on a D-$p=3$
brane world volume. It will prove useful to
 work out most formulas for generic $(D,d)$.
\par
Old-fashioned Chan-Paton factors are transformed in a sense, 
in Dirichlet boundary conditions for open strings, the $U(N)$ gauge theory 
being thus represented by a stack of $N$ parallel D-branes.
The standard lore is to argue that the $U(1)$ part just represents the
center of mass degree of freedom. We shall indeed concentrate in this section
on the nonabelian subgroup $SU(N)$, and leave the abelian part to the next
section.
\par

In the presence of two Dirichlet p-branes, the calculation of the vector
amplitude is modified only in two places: first of all, there is a zero 
mode correction in the integrand, namely:
\be
e^{-\frac{\tau y^2}{2\pi^2 \a'}}
\ee
where $y$ is the separation between the branes, and the rest of the notation
is as before.
\par
The second modification\footnote{Were the external momenta to have 
non-vanishing transversal components, Dirichlet Green functions ought
to be employed.}
 is physically much more important. The dimension $d$
we put before in the general expression (\ref{ampli}) is now replaced
by $d\equiv p+1$, the total worldvolume dimension of the brane,
{\em except} that the string can vibrate in the full target spacetime,
so that the exponent of Dedekind's function must be $D$ (that is $26$
in the critical dimension).
\par

In order to extract the field theory limit of this amplitude, we condider
the simultaneous limit where $\a',y,\t^{-1}\rightarrow 0$, keeping
finite Maldacena's variable, $u\equiv y/\a'$ as well as $\tilde{\t}\equiv
\a'\t$ and $\hat{\n}\equiv \n/\t$. 
\par 
It is worth remarking that the variables $\tilde{\t}$ and $u$ are 
somewhat similar geometrically, owing to the fact that $y$ and $\t^{-1}$
are respectively the world-sheet and space-time length of the cylinder.
\par
To be specific, the {\em on-shell} amplitude now reads:
\be\label{A}
A= \frac{g^2}{(4\pi)^{d/2}}\epsilon_1.\epsilon_2
\int\frac{d\bar{\t}}{\bar{\t}}\int_0^1 d\hat{\nu}\bar{\t}^{1-d/2}
e^{- u^2 \bar{\t}}[(D-2)(1- 2 \bar{\nu})^2 - 8]
\ee

From where we easily get:
\be
A= -\frac{g^2}{(4\pi)^{d/2}}\epsilon_1.\epsilon_2
\Gamma(1- d/2) (u^2)^{d/2 -1} [(D-2) - 4 (D-2) B(2,2) - 8]
\ee
(where $B(u,v)$ is Euler's Beta function),
which by standard dimensional regularization leads for $d=4$ to:
\be\label{u}
A= -\frac{g^2}{(4\pi)^{d/2}}\epsilon_1.\epsilon_2
u^2 ~log(u^2/\Lambda^2)[(D-2) - 4(D-2) B(2,2) - 8]
\ee

It has been already pointed out that Maldacena's variable $u$ plays naturally
the r\^ole of the renormalization group scale (cf. \cite{alvarezgomez}).
\par

We notice finally, that, in contrast with (\ref{ampli}), D-brane 
amplitudes can perfectly well be considered on shell, with external momenta
parallel to the D-brane, and thus with only $d\equiv p+1$ 
non trivial components but, we insist, with oscillatory modes of the 
intermediate strings
vibrating in all $D$ dimensions.
\par

A further observation is that when computing the planar amplitude,
the factor $ \frac{6-7d}{d-1}$ in eqn.(\ref{desc})
is now replaced by $\frac{D+6-8d}{d-1}$. This in turn leads to the
new beta function as given by:
\be
\b (g) =  \frac{1}{2}\frac{D+6-8d}{d-1}C_2^{adj}(G) \frac{g^3}{16\pi^2}
\ee
\par
It is worth noticing that $\b = 0$ when $D=26$ and $d=4$, which corresponds
to Yang-Mills with $n_s=22$ scalar flavours in the adjoint of $SU(N)$
(\cite{fradkin}).
Incidentaly, this theory is known {\em not} to be conformal at two loops
(the beta function is instead positive). It would be exceedingly interesting
to perform this calculation in a stringy context, to check whether
the quantum field theory contribution to the two-loop beta function
can be related to contributions coming from the region of the genus two
string diagram in which closed string states are exchanged.

\section{Abelian groups in the infinite tension limit}
To start with we will reduce ourselves in this section to repeat
our previous analysis for abelian gauge groups. What we should
naturally expect as infinite tension field theory limit would be 
pure Maxwell theory without charged matter and therefore trivially vanishing
beta function. This is what seems to indicate the field theory
limit of tree level string amplitudes with abelian Chan-Paton factors, 
where for instance the three point amplitude vanishes as a consequence
of cyclic invariance. When we try to check the validity of this 
picture going to one loop amplitudes we get some surprises that could
be potentially interesting to understand the low energy field theory 
on the world volume of just one D-brane.
\par
Coming back to the analysis of the two point amplitude 
nothing much changes in the Abelian case, except for the fact that
the nonplanar diagram where the vertex 
operators are inserted in different boundaries does not vanish anymore.
This means that 
the situation is in some sense, the opposite as in field theory:
now we have two diagrams instead of only one. The non planar diagram in 
the full fleged string 
corresponds to the well known reggeon-pomeron vertex; a vertex where the
open string without Chan-Paton factors becomes a closed string. In the
limit $\t=0$ the corresponding diagram  is dominated by light closed string
states that can mix with the photon giving rise to a stringy version of Higgs
mechanism known as Cremmer Scherk \cite{cremmer} phenomena. In this paper
we will not be interested in this limit ( that away from the critical dimension
produces cuts violating unitarity ) but instead in the $\t=\infty$ limit
of the non planar diagram.
\par
In order to capture the essential difference between abelian and 
non-abelian, let us also perform this computation in detail.
\par
The first part of the calculation is already done; the planar
diagram is {\em exactly} the same as before; with the only difference
that the factor coming from the external traces is now $(\frac{1}{\sqrt{2}})^2
$ 
instead of
$N/2 \delta_{ab}$.\footnote{This is because if the $U(1)$ is considered
as imbedded in $U(N)$, it is natural to keep the normalization of
 the generators in such a way that $tr ~T^2 = \frac{1}{2}$.}
\par
The nonplanar diagram can be similarly computed as:
\bea
&& C_{NP}(p_1,p_2)= \frac{1}{2}\frac{g^2 \mu^{4-d}}{(4\pi)^{d/2}}
(2\a')^{1-d/2}
\nonumber\\
&&\int_{0}^{\infty}d\tau \int_{0}^{s}d\nu \tau^{-d/2} 
e^{\frac{(26-d)\tau}{12}}
\eta(i\tau/\pi)^{-(d-2)}\nonumber\\
&&[-(\epsilon_1.\epsilon_2) \pd_{\nu_1}\pd_{\nu_{2}}G_{P}(\nu_1,\nu_2) + 
\a' (\epsilon_1.p_2)(\epsilon_2.p_1)\pd_{\nu_1}
G_{NP}(\nu_1,\nu_2)\pd_{\nu_2}G_{NP}(\nu_1,\nu_2)]\nonumber\\
&&e^{2\a' p_1.p_2 G_{NP}(\nu_1,\nu_2)}
\eea
$G_{NP}(\nu_1,\nu_2)$ is the Green function computed between different
boundaries (that is, $G_{NP}\equiv 
G(i\frac{\nu}{\pi}+\frac{1}{2}|i\frac{\tau}{\pi})$)\footnote{
There is an extra factor of $i$ coming from the zero modes, cf. \cite{green}}
, namely:
\be
G_{NP}(\nu_1,\nu_2)= \log~-2\pi ~\frac{\theta_2(\frac{i}{\pi}(\nu_2-\nu_1)| 
\frac{i \tau}{\pi})}{\theta'_{1}(0|\frac{i\tau}{\pi})} - 
\frac{(\nu_2 - \nu_1)^2}{\tau} 
\ee
The expansion of the Green function is:
\be
G_{NP}\sim -\frac{\nu^2}{\tau}+\n +\log~(1+e^{-2\n})+ e^{-2\tau}[\frac{e^{-3\n}
+e^{3\n}}{e^{\n}+e^{-\n}}+3]
\ee
which yields for the product of the derivatives of the Green functions:
\be
\pd G_{NP}\pd G_{P}\sim (d-2)(1-2\hat{\n})^2 + 8 ~e^{-2 \tau}
\ee
The last term cancels when added to the planar diagram, and the final result
for the (divergent part of) the sum of the two contributions is:
\be\label{np}
C_{NP}(s) = 2~\frac{1}{2}~\frac{g^2}{(4\pi)^2}~\frac{d-2}{d-1}
~\Gamma(\frac{4-d}{2})
s \e_1.\e_2 B(1,1) 
\ee
leading to\footnote{Actually the counting is very simple: the non surviving
part of the factor $11/3$ is the non UV free, that is, $1/3$ and this 
appears twice, once for each string diagram} :
\be
\b = 2~\frac{1}{48\pi^2}g^3
\ee
This is {\em exactly} the correct result for scalar QED with two $= (d-2)$
flavours in the fundamental representation of the abelian gauge group.!
\par 
It comes, however, as 
a surprise, because if we only had gauge particles in this limit, it
being a free theory, we should expect $\beta = 0$.
\par
The origin of the non trivial beta function goes back to the structure
of string amplitudes where the Chan-Paton group labels appears as
multiplicative factors. So for instance the planar diagram is , up
to a numerical factor, the same for abelian and non abelian groups. The
specific abelian contribution corresponding to the non planar diagram
is not cancelling the planar amplitude since the zero mode part of $G$ and 
$G_{NP}$ are the same. What cancels between the planar and the non planar
amplitudes is the piece coming from what we denote amplifying tachyon effect.
Thus the field theory limit of string amplitudes with abelian Chan-Paton
factors is not pure Maxwell but instead scalar QED, more precisely the
scalar QED  describing the diamagnetic part of the Yang Mills beta function.
\par
If we repeat the previous analysis using a D-brane picture we
get a beta function depending on the target space time dimension $D$ of the
type:
\be
\b(g) = \frac{D-2}{d-1} \frac{1}{16\pi^2} g^3
\ee
which for $d=4$ is that of QED with $(D-2)$ charged scalar particles.
It is important to notice that the number of scalars contributing
to the beta function is independent of the D-brane world volume dimension 
$d$. Notice also that contrary to what did happen for the non abelian 
case where
we found $\beta=0$ for $D=26$ here we get $\beta=0$ for $D=2$ that curiously
enough is the critical dimension for the $U(1)$ string model of 
\cite{ademollo}.
\par
The previous analysis of $U(1)$ amplitudes lead us naturally to
the conclusion that the low energy field theory description of one D-3 brane
is scalar QED with $(D-2)$ charged scalars. The physics underlying
these extra scalars can be due to the dual interpretation of loop corrections
in perturbation theory as tree level amplitudes where one of the states is
a closed string state i.e part of the gravitational sector. It could be
this interaction with gravity the one resposible
for the non vanishing beta function and therefore for
the unexpected new degrees of freedom.

\section*{Acknowledgments}
We are grateful to Ian Kogan for inspiring discussions.
This work ~~has been partially supported by the
European Union TMR program FMRX-CT96-0012 {\sl Integrability,
  Non-perturbative Effects, and Symmetry in Quantum Field Theory} and
by the Spanish grant AEN96-1655.  The work of E.A.~has also been
supported by the European Union TMR program ERBFMRX-CT96-0090 {\sl 
Beyond the Standard model} 
 and  the Spanish grant  AEN96-1664.





\begin{thebibliography}{99}



\bibitem{abbott} L.F. Abbott,
                {\em The Background Field Method Beyond One Loop},
                 Nucl. Phys. B185(1981),189.\\
                L.F. Abbott, M.T. Grisaru and R. K. Schaeffer,
                {\em The Background Field Method and the S-Matrix},
                 Nucl.Phys. B229,(1983),372.







\bibitem{ademollo}M. Ademollo et al., 
                  {\em Dual String with U(1) color symmetry},
                  Nucl.Phys.B111:77-110,1976.






\bibitem{alvarezgomez}E. Alvarez and C. G\'omez,
                  {\em Noncritical confining strings and the
                   renormalization group},
                  {\tt hep-th 9902012}; Nucl. Phys. B550 (1999),169.






\bibitem{alvarezgaume} L. Alvarez Gaume, C. Gomez, G. Moore, C. Vafa, 
                       {\em Strings in the Operator Formalism},
                       Nucl.Phys.B303:455,1988.








\bibitem{bern}   Zvi Bern,
                 {\em String-based Perturbative Methods 
                 for Gauge Theories},
                 {\tt hep-th/9304249}.
                 (Presented at Theoretical Advanced Study Institute 
                 (TASI 92): From Black Holes and Strings to Particles,
                 Boulder, CO, 3-28 Jun 1992). 
                 Published in Boulder TASI 92:0471-536 






\bibitem{burgess} C.P. Burgess and T.R. Morris,
                 {\em Open and Unoriented Strings \`a la Polyakov},
                 Nucl. Phys. B291(1987)256.







\bibitem{cremmer} E. Cremmer and J. Scherk,
                  {\em Spontaneous dynamical breaking of gauge symmetry
                  in dual models},
                  Nucl. Phys. B72 (1974) 117.




\bibitem{divecchia} P. Di Vecchia, L. Magnea, A. Lerda, R. Russo 
                    and R. Marotta,
                    {\em String Techniques 
                    for the Calculation of Renormalization 
                    Constants in Field Theory},
                    Nucl. Phys. B469 (1996) 235; {\tt hep-th/9601143}



\bibitem{fradkin} E.S. Fradkin and A. Tseytlin, 
                  {\em Quantum properties of higher dimensional and
                  dimensionally reduced supersymmetric theories}, 
                  Nucl.Phys. B227 (1983) 252.


 

\bibitem{green} M.B.Green.J.H.Schwarz and E. Witten,
                {\em Superstring Theory}
                 (Cambridge University Press,1987)




\bibitem{gross} D.J. Gross and F. Wilczek,
                Phys. Rev. Lett. 30 (1973)1343;\\
                H.D. Politzer,
                Phys. Rev. Lett, 30 (1973)1346.







\bibitem{metsaev}R. Metsaev and A. Teytlin, 
                 {\em On loop corrections to string theory effective actions},
                 Nucl.Phys. B298 (1988) 109,



\bibitem{minahan} J.A. Minahan,
                  {\em One-Loop Amplitudes on Orbifolds and the
                  Renormalization of Coupling Constants},
                   Nucl. Phys. B298 (1988),36.



\bibitem{nielsen} N.K. Nielsen,
                   {\em Asymptotic freedom as a spin effect},
                   Am. J. Phys.49 (1981),1171.





\end{thebibliography}
\end{document}